----------------------------------------------------------
\documentstyle[aps,12pt]{revtex}
\begin{document}

\title{Quantum four-body system in $D$ dimensions}

\author{Xiao-Yan Gu \thanks{Electronic address:
guxy@mail.ihep.ac.cn} and Zhong-Qi Ma \thanks{Electronic address:
mazq@sun.ihep.ac.cn}}

\address{CCAST (World Laboratory), P.O.Box 8730, Beijing 100080, China \\
and Institute of High Energy Physics, Beijing 100039, China}

\author{Jian-Qiang Sun \thanks{Electronic address:
sunjq@mail.ihep.ac.cn}}

\address{Institute of High Energy Physics, Beijing 100039, China}

\maketitle

\date{}

\vspace{5mm}
\begin{abstract}
By the method of generalized spherical harmonic polynomials,
the Schr\"{o}dinger equation for a four-body system in
$D$-dimensional space is reduced to the generalized radial
equations where only six internal variables are involved. The
problem on separating the rotational degrees of freedom
from the internal ones for a quantum $N$-body system in $D$
dimensions is generally discussed.

\end{abstract}

\section{Introduction}

Recent years have witnessed a flurry of investigations into the
arbitrary $D$-dimensional problems \cite{ben,rom,mav,kha,kir} in
many branches of physical chemistry and chemical physics. The
problems associated with the $D$-dimensional hydrogen atom
\cite{kos,ban,nie}, the $D$-dimensional harmonic oscillator
\cite{lou,sha,bak}, and the connection between the two
\cite{fiv,aeb,goe,kib,nou} have been thoroughly discussed. During
the past few years, with the application of dimensional scaling to
the quantum theory of atomic and molecular structure, large-$D$
helium problem has also been studied by many authors
\cite{dup,dor,her,goo,cha}. This approach requires solving the
few-body Schr\"{o}dinger equation in a $D$-dimensional coordinate
space and has been applied to a large number of physically
interesting problems \cite{her2,loe,goo2,kai}. Due to the
complexity of the problem for an $N$-body system in $D$
dimensions, so far there is no complete theoretical solution when
$N>3$.

In our recent work \cite{gu1}, a new method for separating
the rotational degrees of freedom from the internal ones in
a few-body system was proposed. The power of this new
approach is in its ability of great simplification in
calculation of energy levels of a few-body system in terms of
the generalized radial equations involved only internal
variables, which are derived from the Schr\"{o}dinger equation
without any approximation. Some typical three-body system in
three-dimensional space, such as a helium atom \cite{duan1,duan2,duan3}
and a positronium negative ion \cite{duan4} have been solved
numerically with high precision. The key to the approach
is that we have found a complete set of independent eigenfunctions
of angular momentum for the system, which are homogeneous
polynomials in the components of Jacobi coordinate vectors
and satisfy the Laplace equation, and chosen a suitable set
of internal variables. Any wave function with a given angular
momentum can be expanded with respect to the base functions
where the coefficients, called the generalized radial functions,
depend only upon the internal variables. The generalized radial
equations satisfied by the generalized radial functions are
easily derived owing to the nice property of the base functions
\cite{gu1}. This method has been generalized to the arbitrary
dimensional space for a three-body system \cite{gu2}. The
exact interdimensional degeneracies in the system can be
obtained directly from the generalized radial equations \cite{gu3}.

To further this study, we expect to apply this approach to
an $N$-body system in $D$ dimensions. As noticed in our
previous paper \cite{gu2}, the cases with $N<D$ are very
different to the cases with $N\geq D$. The general formulas
are hard to express uniformly due to arbitrariness of $D$
and $N$. However, the main characters are manifested fully
in a four-body system of $D$ dimensions, but not fully in a
three-body system. The four-body problems also
play a fundamental role in nuclear and hypernuclear physics
\cite{ric,fon,kam}. In this paper we will study the problem
of separating the rotational degrees of freedom from the
internal ones for a quantum four-body system in $D$
dimensions in some detail. The general case ($N$-body
system) will be summarized.

The plan of this paper is as follows. In Sec. II, after
separating the motion of the center of mass by Jacobi coordinate
vectors, we will define the generalized spherical harmonic
polynomials for a four-body system in $D$ dimensions and prove
that they constitute a complete set of independent base functions
for a given total orbital angular momentum in the system. Some
new features in comparison with the three-body case are also
discussed in this section. The generalized radial equations
satisfied by the generalized radial functions are established
in Sec. III. In Sec. IV, we will generalize this method to
separate the rotational degrees of freedom from the internal
ones for an $N$-body system in $D$ dimensions. Some conclusions
will be given in Sec. V.

\section{The generalized spherical harmonic polynomials}

For a quantum $N$-body system in an arbitrary $D$-dimensional
space, we denote the position vectors and the masses of $N$
particles by ${\bf r}_{k}$ and by $m_{k}$, $k=1,~2,\ldots,~N$,
respectively. $M=\sum_{k} m_{k}$ is the total mass. The
Schr\"{o}dinger equation for the $N$-body system with a pair
potential $V$, depending upon the distance of each pair of
particles, $|{\bf r}_{j}-{\bf r}_{k}|$, is
$$- \displaystyle {1 \over 2}  \bigtriangledown^{2} \Psi
+V \Psi =E \Psi ,~~~~~~ \bigtriangledown^{2}=
 \displaystyle \sum_{k=1}^{N}~
\displaystyle m_{k}^{-1} \bigtriangledown^{2}_{{\bf r}_{k}}, 
 \eqno (1) $$

\noindent where $\bigtriangledown^{2}_{{\bf r}_{k}}$ is the
Laplace operator with respect to the position vector ${\bf
r}_{k}$. For simplicity, the natural units $\hbar=c=1$ are
employed throughout this paper. The total orbital angular momentum
operators $L_{ab}$ in $D$ dimensions are defined as
\cite{cha,lou2}
$$ L_{ab}=-L_{ba}=-i\displaystyle \sum_{k=1}^{N}~\left\{ r_{ka}
\displaystyle {\partial \over \partial r_{kb}} -r_{kb}\displaystyle
{\partial \over \partial r_{ka}} \right\},~~~~~~ a,b=1,2,\ldots D,
\eqno (2) $$

\noindent
where $r_{ka}$ denotes the $a$th component of the position
vector ${\bf r}_{k}$.  Now, we replace the position vectors
${\bf r}_{k}$ with the Jacobi coordinate vectors ${\bf R}_{j}$:
$$ {\bf R}_{0}=M^{-1/2}\displaystyle \sum_{k=1}^{N}~m_{k}
{\bf r}_{k},~~~~~
{\bf R}_{j}=\left(\displaystyle {m_{j+1} M_{j}\over M_{j+1}}  \right)^{1/2}
\left({\bf r}_{j+1}-\displaystyle \sum_{k=1}^{j}~ \displaystyle
{m_{k}{\bf r}_{k}\over M_{j}}\right), $$
$$1\leq j \leq (N-1),~~~~~~M_{j}
=\displaystyle \sum_{k=1}^{j}~m_{k},~~~~~~M_{N}=M,  \eqno (3) $$

\noindent
where ${\bf R}_{0}$ describes the position of the center
of mass, ${\bf R}_{1}$ describes the mass-weighted separation from
the second particle to the first particle. ${\bf R}_{2}$ describes
the mass-weighted separation from the third particle to the center
of mass of the first two particles, and so on. It is
straightforward to illustrate that the potential $V$ is a function
of ${\bf R}_{j}\cdot {\bf R}_{k}$ and is rotationally invariant.

In the center-of-mass frame, ${\bf R}_{0}=0$. A straightforward
calculation by replacement of variables shows that the Laplace
operator in Eq. (1) and the total orbital angular momentum
operator $L_{ab}$ in Eq. (2) are directly expressed in ${\bf R}_{j}$:
$$\begin{array}{l}
\bigtriangledown^{2}=\displaystyle \sum_{j=1}^{N-1}~
\bigtriangledown^{2}_{{\bf R}_{j}},~~~~~~
L_{ab}=\displaystyle \sum_{j=1}^{N-1}~L^{(j)}_{ab}
=-i\displaystyle \sum_{j=1}^{N-1}~\left\{
R_{ja}\displaystyle {\partial \over \partial R_{jb}}
-R_{jb}\displaystyle {\partial \over \partial R_{ja}} \right\},\\
{\bf L}^{2}=\displaystyle \sum_{a<b=2}^{D}~L_{ab}^{2},~~~~~~
\left({\bf L}^{(j)}\right)^{2}=\displaystyle \sum_{a<b=2}^{D}~
\left(L_{ab}^{(j)}\right)^{2}. \end{array} \eqno (4) $$

For a four-body system, there are three Jacobi coordinate
vectors ${\bf R}_{1}$, ${\bf R}_{2}$ and ${\bf R}_{3}$,
which will be denoted for simplicity by ${\bf x}$, ${\bf y}$
and ${\bf z}$, respectively:
$$\begin{array}{rl}
{\bf x}&=~\left[\displaystyle {m_{1}m_{2}
\over m_{1}+m_{2}}\right]^{1/2} \left\{{\bf r}_{2}-{\bf r}_{1}
\right\},\\[2mm]
{\bf y}&=~\left[\displaystyle {(m_{1}+m_{2})m_{3}
\over m_{1}+m_{2}+m_{3}} \right]^{1/2}\left\{{\bf r}_{3}
-\displaystyle {m_{1}{\bf r}_{1}+m_{2}{\bf r}_{2}\over
m_{1}+m_{2}}\right\}, \\[2mm]
{\bf z}&=~\left[\displaystyle {(m_{1}+m_{2}+m_{3}+m_{4})m_{4} \over
m_{1}+m_{2}+m_{3}} \right]^{1/2}{\bf r}_{4}. \end{array} \eqno (5) $$

\noindent
Hence,
$$\begin{array}{ll}
\bigtriangledown^{2}=\bigtriangledown^{2}_{\bf x}
+\bigtriangledown^{2}_{\bf y}+\bigtriangledown^{2}_{\bf z},~~~~~~
&L_{ab}=L^{(x)}_{ab}+L^{(y)}_{ab}+L^{(z)}_{ab}\\
{\bf L}^{2}=\displaystyle \sum_{a<b=2}^{D}~L_{ab}^{2},~~~~~~
&\left[{\bf L}^{(x)}\right]^{2}=\displaystyle \sum_{a<b=2}^{D}~
\left[L_{ab}^{(x)}\right]^{2},\\
\left[{\bf L}^{(y)}\right]^{2}=\displaystyle \sum_{a<b=2}^{D}~
\left[L_{ab}^{(y)}\right]^{2},~~~~~~
&\left[{\bf L}^{(z)}\right]^{2}=\displaystyle \sum_{a<b=2}^{D}~
\left[L_{ab}^{(z)}\right]^{2}. \end{array} \eqno (6) $$

\noindent
The Schr\"{o}dinger equation (1) for $D\geq N=4$ reduces to
$$\begin{array}{l}
\left\{\bigtriangledown^{2}_{\bf x}
+\bigtriangledown^{2}_{\bf y}+\bigtriangledown^{2}_{\bf z}
\right\}\Psi({\bf x,y,z})
=-2\left\{E-V\left(\xi_{j}, \eta_{j},\zeta_{j}\right)
\right\}\Psi({\bf x,y,z}),\\
\xi_{1}={\bf x}\cdot {\bf x},~~~~~
\xi_{2}=\eta_{1}={\bf x}\cdot {\bf y},~~~~~
\xi_{3}=\zeta_{1}={\bf x}\cdot {\bf z},\\
\eta_{2}={\bf y}\cdot {\bf y},~~~~~
\eta_{3}=\zeta_{2}={\bf y}\cdot {\bf z},~~~~~
\zeta_{3}={\bf z}\cdot {\bf z}.
\end{array} \eqno (7) $$

\noindent where
$\xi_{j}$, $\eta_{j}$, and $\zeta_{j}$ are internal variables.
It is worth noticing that for the cases $3=D<N$ two Jacobi 
coordinate vectors ${\bf x}$ and ${\bf y}$ can determine 
the body-fixed frame and this set of internal variables 
is not complete because two configurations with different 
directions of ${\bf z}$ reflecting to the plane spanned by 
${\bf x}$ and ${\bf y}$ are described by the same internal 
variables. As pointed in Ref. \cite{gu1}, the variables $\zeta_{3}$
has to be changed to $\left({\bf x}\times {\bf y}\right)\cdot {\bf z}$.
We will further discuss this problem in Sec. IV.

Since Eq. (7) is rotational invariant, the total orbital angular
momentum is conserved. As discussed in Ref. \cite{gu2}, in
$D$-dimensional space, the wave function $\Psi({\bf x,y,z})$ with
a given total angular momentum has to belong to an irreducible
representation of SO($D$), and the angular momentum is also
denoted by the representation. For a four-body system, there are
only three Jacobi coordinate vectors so that the possible
irreducible representation is described by a three-row
Young pattern $[\mu,\nu,\tau]$ of SO$(D)$, or its highest
weight ${\bf M}=(M_{1},M_{2},M_{3},0,\ldots,0)$, where
$$M_{1}=\mu-\nu,~~~~~~M_{2}=\nu-\tau,~~~~~~
M_{3}=\tau. \eqno (8) $$

\noindent
We only need to consider the highest weight state
$\Psi^{[\mu,\nu,\tau]}_{\bf M}({\bf x,y,z})$
because its partners can be calculated from it by the lowering
operators. In this paper the highest weight state will be
simply called the wave functions with the given angular
momentum $[\mu,\nu,\tau]$ for simplicity.

Now we are going to find a complete set of independent
eigenfunctions of total orbital angular momentum, where
"independent" means that each one in the set cannot be expressed
as a combination of the remaining with coefficients only depending
on the internal variables. As discussed in our previous paper
\cite{gu2}, the spherical harmonic polynomials ${\cal
Y}^{(l)}_{\bf m}(\hat{\bf x})$ are homogeneous polynomials in the
components of ${\bf x}$ of degree $l$, spanning an irreducible
traceless tensor space describes by the Young pattern $(l)\equiv
[l,0,0]$. When $D>6$, the explicit forms for some polynomials with
higher weights ${\bf m}$ are as follows \cite{gu4}:
$$\begin{array}{l}
{\cal Y}^{(l)}_{(l)}({\bf x})=N_{D,l}(x_{1}+ix_{2})^{l},\\
{\cal Y}^{(l)}_{(l-2,1,0,\ldots,0)}({\bf x})
=-\sqrt{l}N_{D,l}(x_{1}+ix_{2})^{l-1}(x_{3}+ix_{4}),\\
{\cal Y}^{(l)}_{(l-4,2,0,\ldots,0)}({\bf x})
=\sqrt{l(l-1)/2}N_{D,l}(x_{1}+ix_{2})^{l-2}(x_{3}+ix_{4})^{2},
\end{array} \eqno (9) $$

\noindent
where the last equality holds for $l>1$, and $N_{D,l}$
denotes the normalization factor given in \cite{gu4}.
The product of two spherical harmonic polynomials
${\cal Y}^{(l)}_{\bf m}(\hat{\bf x})$
and ${\cal Y}^{(l')}_{\bf m'}(\hat{\bf y})$ belongs to the direct
product of two representation $(l)$ and $(l')$, which is
a reducible representation. It can be reduced by the
Littlewood-Richardson rule and contraction of a pair of
$x_{a}$ and $y_{a}$, where the latter relates to the internal
variables:
$$(l)\otimes (l')\simeq \displaystyle \bigoplus_{s=0}^{min\{l,l'\}}~
\displaystyle \bigoplus_{t=0}^{min\{l,l'\}-s}~
[l+l'-s-2t,s,0]. \eqno (10) $$

\noindent
Since a base function containing a factor depending on internal
variables is not independent, only those representations
$[l+l'-s,s,0]$ [$t=0$ in Eq. (10)] calculated by the
Littlewood-Richardson rule are related to the independent base
functions \cite{gu2}. Calculating by the Clebsch-Gordan coefficients
and removing the normalization factor, we obtain the independent
base functions for the representations $[l+l'-s,s,0]$, called
the generalized spherical harmonic polynomial
$Q^{(l+l'-s) s}_{l}({\bf x,y})$. Changing the parameters
$\mu=l+l'-s$, $\nu=s$ and $q=l$, we define the generalized
spherical harmonic polynomial $Q^{\mu \nu}_{q}({\bf x,y})$
for the representation $[\mu,\nu]$ \cite{gu2} as
$$\begin{array}{l}
Q^{\mu \nu}_{q}({\bf x,y})=\displaystyle {X_{12}^{q-\nu}
Y_{12}^{\mu-q}\over (q-\nu)!(\mu-q)!}
(X_{12}Y_{34}-Y_{12}X_{34})^{\nu},~~~~~~~~0\leq \nu \leq q \leq \mu,\\
X_{12}=x_{1}+ix_{2},~~~~X_{34}=x_{3}+ix_{4},~~~~
Y_{12}=y_{1}+iy_{2},~~~~Y_{34}=y_{3}+iy_{4}. \end{array} \eqno (11) $$

For the product of three spherical harmonic polynomials,
Eq. (10) is generalized to
$$\begin{array}{l}
(l)\otimes (l') \otimes (l") \\
~~~\simeq \displaystyle \bigoplus_{r=0}^{min\{l,l'\}}~~
\displaystyle \bigoplus_{\nu=r}^{min\{(l+l'-r),(r+l")\}}~~
\displaystyle \bigoplus_{\tau=0}^{min\{r,(l"-\nu+r)\}}~
[l+l'+l"-\nu-\tau,\nu,\tau] \oplus \ldots . \end{array}  \eqno (12) $$

\noindent
The ellipsis denotes those representations related to
the base functions which are not independent.

Filling the digits $1$, $2$ and $3$ arbitrarily into a given
Young pattern $[\mu, \nu, \tau]$ ($\mu\geq \nu \geq \tau$) we
obtain a young tableau. A Young tableau is called standard if the
digit in every column of the tableau increases downwards and the
digit in every row does not decrease from left to right. In fact,
the digits $"1"$, $"2"$ and $"3"$ denote the components of ${\bf
x}$, ${\bf y}$, and ${\bf z}$, respectively. Obviously, the
representation $[l+l'+l"-\nu-\tau,\nu,\tau]$ listed in Eq. (12)
corresponds to a standard Young tableau, where the number of
digit $"1"$ in the first row is $l$, the numbers of digit
$"2"$ in the first and the second rows are respectively $(l'-r)$
and $r$, and the numbers of digit $"3"$ in the first, second and
third rows are respectively $(l"+r-\nu-\tau)$, $\nu-r$ and $\tau$.
The base functions in the remaining representation spaces, which
correspond to non-standard Young tableaux, are not independent.

For a given pattern $[\mu,\nu,\tau]$, each standard Young tableau
is determined by three parameters $q$, $p$ and $r$, where $q$ is
the number of digit $"1"$ in the first row, $p$ and $r$ are the
numbers of digit $"2"$ in the first and the second rows,
respectively. $q$, $p$ and $r$ should satisfy the constraints:
$\tau \leq r \leq q$ and $r \leq \nu \leq q+p\leq \mu$. The number
of standard Young tableaux for the given Young pattern $[\mu, \nu,
\tau]$ is equal to the dimension of the representation $[\mu, \nu,
\tau]$ of the SU(3) group:
$$d_{[\mu,\nu, \tau]}(SU(3))=\displaystyle {1\over 2} (\mu-\tau+2)
(\nu-\tau+1)(\mu-\nu+1).  \eqno (13) $$

For a given representation $[\mu,\nu,\tau]$ of SO($D$), each
standard Young tableau denoted by $(q,p,r)$ corresponds to a
representation space. The highest weight state in the representation
space $(q,p,r)$ is the generalized spherical harmonic polynomial
$Q_{q p r}^{\mu \nu \tau}({\bf x, y, z})$:
$$\begin{array}{l}
Q_{q p r}^{\mu \nu \tau}({\bf x, y, z})
=\left\{\begin{array}{ll}
\displaystyle{ X_{12}^{q-\nu}Y_{12}^{p}Z_{12}^{\mu-q-p}
T_{12}^{r-\tau}T_{13}^{\nu-r}T^{\tau} \over   (q-\nu)!p!(\mu-q-p)!
(r-\tau)!(\nu-r)!},~~~&{\rm when}~~q  \geq \nu, \\
\displaystyle{ Y_{12}^{q+p-\nu}Z_{12}^{\mu-q-p}
T_{12}^{r-\tau}T_{13}^{q-r}T_{23}^{\nu-q}T^{\tau} \over
(q+p-\nu)!(\mu-q-p)!(r-\tau)!(q-r)!(\nu-q)!},~~~&{\rm
when}~~q<\nu,
\end{array} \right.\\
\tau \leq r \leq q,~~~~~~~~r \leq \nu \leq q+p \leq \mu, \\
 X_{12}=x_{1}+ix_{2},
~~~~~~ Y_{12}=y_{1}+iy_{2},~~~~~~Z_{12}=z_{1}+iz_{2}, \\
 X_{34}=x_{3}+ix_{4}, ~~~~~~Y_{34}=y_{3}+iy_{4},
~~~~~~ Z_{34}=z_{3}+iz_{4}, \\
X_{56}=x_{5}+ix_{6},~~~~~~Y_{56}=y_{5}+iy_{6},
~~~~~~Z_{56}=z_{5}+iz_{6},\\
T_{12}=X_{12}Y_{34}-X_{34}Y_{12},~~~~T_{13}=X_{12}Z_{34}-X_{34}Z_{12},
~~~~T_{23}=Y_{12}Z_{34}-Y_{34}Z_{12},\\
T=X_{12}Y_{34}Z_{56}+X_{34}Y_{56}Z_{12}+X_{56}Y_{12}Z_{34}-
X_{12}Y_{56}Z_{34}-X_{34}Y_{12}Z_{56}-X_{56}Y_{34}Z_{12}.
 \end{array} \eqno (14) $$

It is evident that $Q_{q p r}^{\mu \nu \tau}({\bf x, y, z})$ do
not contain a function of the internal variables as a factor, nor
do their partners due to the rotational symmetry. Therefore, $Q_{q
p r}^{\mu \nu \tau}({\bf x, y, z})$ are independent base functions
for the given angular momentum described by $[\mu,\nu,\tau]$. Due
to Eq. (12), the set of $Q_{q p r}^{\mu \nu \tau}({\bf x, y, z})$
is complete. The reason why the generalized spherical harmonic
polynomial denoted by a non-standard Young tableau is not
independent can be seen from the following identity:
$$T_{23}X_{12}=T_{13}Y_{12}-T_{12}Z_{12},~~~~~~
\begin{tabular}{|c|c|} \hline
2 & 1 \\ \hline 3 & \multicolumn{1}{c}{ } \\  \cline{1-1}  \end{tabular}
=~\begin{tabular}{|c|c|} \hline
1 & 2 \\ \hline 3 & \multicolumn{1}{c}{ } \\  \cline{1-1}  \end{tabular}
~-~\begin{tabular}{|c|c|} \hline
1 & 3 \\ \hline 2 & \multicolumn{1}{c}{ } \\ \cline{1-1}  \end{tabular}~.
 \eqno (15) $$

\noindent
This identity is similar to the Fock's cyclic symmetry condition
\cite{ham}. The left-hand-side of Eq. (15) corresponds to a
non-standard Young tableau, and two terms in the right-hand-side
correspond to two standard Young tableaux, respectively.

Since the problem on completeness of the set is very important,
we are going to prove this problem by another method. On the one
hand, because the base function
$Q_{q p r}^{\mu \nu \tau}({\bf x, y, z})$ is a homogeneous
polynomial of degree $\mu+\nu+\tau$ in the components of
${\bf x}$, ${\bf y}$ and ${\bf z}$, we calculate the number
$R_{D}(l)$ of base functions in the sets for the representations
$[\mu,\nu,\tau]$ with $\mu+\nu+\tau=l$. Namely, we want to
calculate how many homogeneous polynomials of degree $l$ exist
in the sets of the independent base functions . We first calculate
how many base functions exist in the set for a given
representation $[\mu,\nu,\tau]$. The dimension of the
representation $[\mu,\nu,\tau]$ of SO($D$) is $d_{D}([\mu,\nu,\tau])$:
$$\begin{array}{rl} d_{D}([\mu,\nu,\tau])
&=~(D+2\mu-2)(D+\mu+\nu-3)(D+\mu+\tau-4)(D+2 \nu-4)\\
&~~~\times (D+\nu+\tau-5)(D+2 \tau-6)(\mu-\tau+2)
(\mu-\nu+1)(\nu-\tau+1)\\
&~~~\times
\displaystyle {(D+\mu-5)!(D+\nu-6)!(D+\tau-7)!\over
(D-2)! (D-4)! (D-6)! (\mu+2)! (\nu+1)!  \tau! }.
\end{array} \eqno (16) $$

\noindent Thus, the number of base functions in the set for the
representation $[\mu,\nu,\tau]$ is
$d_{[\mu,\nu,\tau]}(SU(3))d_{D}(\mu,\nu,\tau)$. Then, the number
$R_{D}(l)$ of base functions in the sets for the representation
$[\mu,\nu,\tau]$ with $\mu+\nu+\tau=l$ is:
$$R_{D}(l)=\displaystyle \sum_{\tau=0}^{[l/3]}
\sum_{\nu=\tau}^{[(l-\tau)/2]}~
d_{[(l-\nu-\tau),\nu,\tau]}(SU(3))~d_{D}([(l-\nu-\tau),\nu,\tau]),
\eqno (17) $$

\noindent
where $[x]$ denotes the largest integer less than or equal to $x$.

On the other hand, the number of linear independent homogeneous
polynomials of degree $l$ in the components of ${\bf x}$, ${\bf y}$
and ${\bf z}$ is $M_{D}(l)$:
$$M_{D}(l)=\left(\begin{array}{c} l+3D-1 \\ 3D-1 \end{array} \right) . $$

\noindent
After removing those polynomials in the form $\xi_{j}f({\bf x,y,z})$,
$\eta_{j}f({\bf x,y,z})$ and $\zeta_{j}f({\bf x,y,z})$ where
$f({\bf x,y,z})$ is a polynomial of degree $(l-2)$, the number
$M_{D}(l)$ reduces to $K_{D}(l)$:
$$\begin{array}{rl}
K_{D}(l)&=~M_{D}(l)-6M_{D}(l-2)+15M_{D}(l-4)
-20M_{D}(l-6)\\
&~~~+15M_{D}(l-8)-6M_{D}(l-10)+M_{D}(l-12)\\
&=~\left\{ (3D-7) (3D-8)(3D-9)(3D-10)(3D-11)(3D-12)\right.\\
&~~~+ 12 l (D-4)[72+(3D-10)(3D-11)(27 D^2-153 D+236)]\\
&~~~+ 4 l^2 [184+45 (D-4)(3D-11)(9 D^2-57D+98)] \\
&~~~+ 480 l^3 (D-4)(9 D^{2}-63 D+126)+ 80 l^4 (27D^{2}-207 D+404)\\
&\left.~~~+ 576 l^5 (D-4) + 64 l^6 \right\}
 \displaystyle {(l + 3D - 13)! \over l!(3D - 7)!},
\end{array} \eqno (18) $$

\noindent
where $l+3D\geq 13$ and $K_{4}(0)=1$. It is checked by
MATHEMATICA that
$$R_{D}(l)=K_{D}(l). \eqno (19) $$

\noindent Thus, we have proved again that $d_{[\mu,\nu,
\tau]}(SU(3))$ polynomials $Q_{q p r}^{\mu \nu \tau}({\bf x, y,
z})$ construct a complete set of independent base functions for
the angular momentum $[\mu,\nu,\tau]$.

The generalized spherical harmonic polynomial
$Q_{q p r}^{\mu \nu \tau}({\bf x, y, z})$ is a homogeneous
polynomial of degrees $q$, $(p+r)$ and $(\mu+\nu+\tau-q-p-r)$
in the components of ${\bf x}$, ${\bf y}$ and ${\bf z}$,
respectively. It is a simultaneous eigenfunction of
$\bigtriangledown^{2}_{\bf x}$, $\bigtriangledown^{2}_{\bf y}$,
$\bigtriangledown^{2}_{\bf z}$,
$\bigtriangledown_{\bf x}\cdot\bigtriangledown_{\bf y}$,
$\bigtriangledown_{\bf x}\cdot\bigtriangledown_{\bf z}$,
$\bigtriangledown_{\bf y}\cdot\bigtriangledown_{\bf z}$, and the
angular momentum operators ${\bf L}^{2}$, $\left[{\bf L}^{(x)}\right]^{2}$,
$\left[{\bf L}^{(y)}\right]^{2}$, $\left[{\bf L}^{(z)}\right]^{2}$,
$$\begin{array}{l}
\bigtriangledown^{2}_{\bf x} Q_{q p r}^{\mu \nu \tau}({\bf x, y, z})
=\bigtriangledown^{2}_{\bf y} Q_{q p r}^{\mu \nu\tau}({\bf x, y, z})
=\bigtriangledown^{2}_{\bf z} Q_{q p r}^{\mu \nu \tau}({\bf x, y, z}) =0,\\
\bigtriangledown_{\bf x}\cdot \bigtriangledown_{\bf y}
Q_{q p r}^{\mu \nu \tau}({\bf x, y, z})
=\bigtriangledown_{\bf x}\cdot \bigtriangledown_{\bf z}
Q_{q p r}^{\mu \nu \tau}({\bf x, y, z})
=\bigtriangledown_{\bf y}\cdot \bigtriangledown_{\bf z}
Q_{q p r}^{\mu \nu \tau}({\bf x, y, z}) =0,\\
{\bf L}^{2}Q_{q p r}^{\mu \nu \tau}({\bf x, y, z})
=C_{2}([\mu, \nu, \tau]) Q_{q p r}^{\mu \nu \tau}({\bf x, y, z}),\\
C_{2}([\mu, \nu,\tau])=\mu(\mu+D-2)+\nu(\nu+D-4)+\tau(\tau+D-6),\\
\left[{\bf L}^{(x)}\right]^{2}Q_{q p r}^{\mu \nu \tau}({\bf x, y, z})
=q(q+D-2) Q_{q p r}^{\mu \nu \tau}({\bf x, y, z}),\\
\left[{\bf L}^{(y)}\right]^{2}Q_{q p r}^{\mu \nu \tau}({\bf x, y, z})
=(p+r)(p+r+D-2) Q_{q p r}^{\mu \nu \tau}({\bf x, y, z}),\\
\left[{\bf L}^{(z)}\right]^{2}Q_{q p r}^{\mu \nu \tau}({\bf x, y, z})
=(\mu+\nu+\tau-q-p-r)(\mu+\nu+\tau-q-p-r+D-2)\\
~~~~~~~~~~~~~~~~~~~~~~~~~~~~~~~~~
\times Q_{q p r}^{\mu \nu \tau}({\bf x, y, z}).
\end{array} \eqno (20) $$

\noindent
where $C_{2}([\mu ,\nu])$ is the Casimir calculated by a
general formula (see (1.131) in Ref. \cite{ma2}). The
parity of $Q_{q p r}^{\mu \nu \tau}({\bf x, y, z}) $ is
obviously equal to $(-1)^{\mu+\nu+\tau}$.

Now, we turn to discuss the case $D\leq 6$. As is well known, the
irreducible traceless tensor space of SO($D$) described by a Young
pattern has the following properties. It is a null space if sum of
the lengths of the first two columns of the Young pattern is
larger than $D$. It reduces into the selfdual and antiselfdual
tensor spaces if the row number of the Young pattern is equal to
$D/2$. Two representations are equivalent if their Young patterns
are the same as each other except for the first column and the sum
of their row numbers is equal to $D$. Those properties cause the
situation for $D\leq 6$ different to that for $D>6$.

When $D=6$, there is no problem for the representation
$[\mu,\nu,\tau]$ with $\tau=0$, but when $\tau\neq 0$,
the representation is reducible. We denote the generalized
spherical harmonic polynomials for the selfdual and
antiselfdual representations by
$Q_{q p r}^{(S)\mu \nu \tau}({\bf x, y, z})$ and
$Q_{q p r}^{(A)\mu \nu \tau}({\bf x, y, z})$, respectively.
$Q_{q p r}^{(S)\mu \nu \tau}({\bf x, y, z})$ is the same as
that given in Eq. (14), and
$Q_{q p r}^{(A)\mu \nu \tau}({\bf x, y, z})$ can be obtained
from $Q_{q p r}^{(S)\mu \nu \tau}({\bf x, y, z})$ by
replacing $X_{56}$, $Y_{56}$ and $Z_{56}$ with $X_{56}^{\prime}$,
$Y_{56}^{\prime}$ and $Z_{56}^{\prime}$:
$$X_{56}^{\prime}=x_{5}-ix_{6},~~~~~~Y_{56}^{\prime}=y_{5}-iy_{6},
~~~~~~Z_{56}^{\prime}=z_{5}-iz_{6}.  \eqno (21) $$

\noindent
The formula (16) for the dimension of the representation
$[\mu,\nu,\tau]$ of SO($D$) holds for $D=6$ when $\tau=0$.
When $\tau\neq 0$, $d_{D}([\mu,\nu,\tau])$ in Eq. (16) is
equal to the sum of the dimensions of the selfdual and
antiselfdual representations such that the equality (19)
still holds for $D=6$.

When $D=5$, in the possible Young pattern $[\mu,\nu,\tau]$,
$\tau$ has to be 0 or 1. The representation $[\mu,\nu,1]$
is equivalent to the representation $[\mu,\nu,0]$. Their
dimensions calculated from Eq. (16) are also the same.
The generalized spherical harmonic polynomials
$Q_{q p r}^{\mu \nu \tau}({\bf x, y, z})$ given in Eq. (14)
hold for $D=5$ except for $x_{6}=y_{6}=z_{6}=0$ and $\tau=0$ or $1$.
Therefore, the equality (19) holds for $D=5$.

For $D=3$, two Jacobi coordinate vectors, say ${\bf x}$ and ${\bf y}$,
can completely determine the body-fixed frame so that the variables
$\zeta_{3}$ has to be changed as
$\left({\bf x}\times {\bf y}\right)\cdot {\bf z}$ in order to
distinguish two configurations with different directions of ${\bf z}$.
We have discussed in detail the four-body system in three dimensions
in our previous paper \cite{gu1}.

The case of $D=4$ is quite complicated because $SO(4)$ is not
a simple group. The representation $[\mu, \nu, 0]$ reduces to
a direct sum of a selfdual representation $[(S) \mu, \nu, 0]$
and an antiselfdual one $[(A) \mu, \nu, 0]$. The generalized
spherical harmonic polynomials $Q_{q p r}^{(S) \mu \nu 0}({\bf x, y, z})$
for the selfdual representations is the same as
$Q_{q p r}^{\mu \nu \tau}({\bf x, y, z})$ with $\tau=0$ given in
Eq. (14). The generalized spherical harmonic polynomials
$Q_{q p r}^{(A) \mu \nu 0}({\bf x, y, z})$ for the antiselfdual
representation can be obtained from
$Q_{q p r}^{(S)\mu \nu 0}({\bf x, y, z})$ by replacing
$X_{34}$, $Y_{34}$ and $Z_{34}$ with $X_{34}^{\prime}$,
$Y_{34}^{\prime}$ and $Z_{34}^{\prime}$:
$$X_{34}^{\prime}=x_{3}-ix_{4},~~~~~~Y_{34}^{\prime}=y_{3}-iy_{4},
~~~~~~Z_{34}^{\prime}=z_{3}-iz_{4}.  \eqno (22) $$

\noindent
If $\tau=1$, then $\nu=1$ and the representation $[\mu,1,1]$
is equivalent to the representation $[\mu,0,0]$. The standard
Young tableau is described by the parameters $q$ and $p$
($r=1$), where $q$ and $p$ are respectively the numbers of
digits $"1"$ and $"2"$ in the first row of the Young
tableau. The generalized spherical harmonic polynomials
for two representations $[\mu,\lambda,\lambda]$, $\lambda=0$,
or 1, are
$$\begin{array}{l}
Q_{q p}^{\mu \lambda \lambda}({\bf x, y, z}) =
\displaystyle{ X_{12}^{q-\lambda}Y_{12}^{p}Z_{12}^{\mu-q-p}
T^{\lambda} \over (q-\lambda)!p!(\mu -q-p)!}, \\
T=X_{12}Y_{34}Z_{34}^{\prime}+X_{34}Y_{34}^{\prime}Z_{12}
+X_{34}^{\prime}Y_{12}Z_{34}-X_{12}Y_{34}^{\prime}Z_{34}
-X_{34}Y_{12}Z_{34}^{\prime}-X_{34}^{\prime}Y_{34}Z_{12}.
 \end{array} \eqno (23) $$

The surprising thing is that Eq. (19) does not satisfied
for $D=4$ and $l\geq 6$. For example,
$$\begin{array}{lccccc}
l=~~&6 & 7 & 8 & 9 & 10 \\
R_{4}(l)~~ & ~5346~ & ~10908~ & ~20550~ & ~36332~ & ~60996~ \\
K_{4}(l)~~ & ~5336 & 10836 & 20256 & 35436 & 58728 \end{array} \eqno (24) $$

\noindent
The reason is that the formula (18) for $K_{D}(l)$ does not hold
for $D=4$ and $l\geq 6$. For $D=4$ we find an identity
with respect to the polynomials of degree $6$ checked
by MATHEMATICA:
$$\begin{array}{l}
\xi_{1}T_{23}^{2}+\eta_{2}T_{13}^{2}+\zeta_{3}T_{12}^{2}
-2\xi_{2}T_{13}T_{23}+2\xi_{3}T_{12}T_{23}-2\eta_{3}T_{12}T_{13}=0.
\end{array} \eqno (25) $$

\noindent
The identity obtained from Eq. (25) by replacing $X_{34}$, $Y_{34}$
and $Z_{34}$ respectively with $X_{34}^{\prime}$, $Y_{34}^{\prime}$
and $Z_{34}^{\prime}$ still holds. Those equalities obtained by
applying the lowering operators and (or) by multiplying a factor to
above two identities are also identities. Thus, the forms
$\xi_{j}f({\bf x,y,z})$, $\eta_{j}f({\bf x,y,z})$ and
$\zeta_{j}f({\bf x,y,z})$, where $f({\bf x,y,z})$ is a homogeneous
polynomial of ${\bf x}$, ${\bf y}$ and ${\bf z}$ of degree $(l-2)$, are
not independent when $l\geq 6$. It is easy to count by
MATHEMATICA that the revised $K_{4}(l)$ by considering
the identities coincides with $R_{4}(l)$.

\section{Generalized radial equations}

In the preceding section we proved that $d_{[\mu,\nu,
\tau]}(SU(3))$ polynomials $Q_{q p r}^{\mu \nu \tau}({\bf x, y,
z})$ construct a complete set of independent base functions for
the angular momentum $[\mu,\nu,\tau]$. Thus, any function
$\Psi^{[\mu,\nu,\tau]}_{\bf M}({\bf x,y,z})$ with angular momentum
$[\mu,\nu,\tau]$ in the system can be expanded with respect to the
base functions $Q_{q p r}^{\mu \nu \tau}({\bf x, y, z})$, where
the coefficients are functions of internal variables.
$$\Psi^{[\mu,\nu,\tau]}_{\bf M}({\bf x,y,z})
=\displaystyle
\sum_{q=\tau}^{\mu} \sum_{p=\max\{(\nu-q),0\}}^{\mu-q} 
\sum_{r=\tau}^{min\{q,\nu\}}~
\psi^{\mu \nu \tau}_{q p r}(\xi_{j},\eta_{j},\zeta_{j})
Q_{q p r}^{\mu \nu \tau}({\bf x, y, z}), \eqno (26) $$

\noindent
where the coefficients
$\psi^{\mu \nu \tau}_{q p r}(\xi_{j},\eta_{j},\zeta_{j})$
are called the generalized radial functions. When substituting
Eq. (26) into the Schr\"{o}dinger equation (5), the main
calculation is to apply the Laplace operator (4) to the
function $\Psi^{[\mu, \nu, \tau]}_{\bf M}({\bf x,y,z})$.
The calculation consists of three parts. In the following, 
we remove the arguments $(\xi_{j},\eta_{j},\zeta_{j})$
and $({\bf x, y, z})$ for simplicity. The first is to apply
the Laplace operator to the generalized radial functions
$\psi^{\mu \nu \tau}_{q p r}(\xi_{j},\eta_{j},\zeta_{j})$,
which can be calculated by replacement of variables: 
$$\begin{array}{rl}
\bigtriangledown^{2} \psi^{\mu \nu \tau}_{q p r}
&=~ \left\{ 4\xi_{1}\partial^{2}_{\xi_{1}} +4\eta_{2}\partial^{2}_{\eta_{2}}
+4\zeta_{3}\partial^{2}_{\zeta_{3}} +2 D\left(\partial_{\xi_{1}}
+\partial_{\eta_{2}} +\partial_{\zeta_{3}}\right)
+\left(\xi_{1}+\eta_{2}\right)\partial^{2}_{\xi_{2}}\right.\\
&~~~+\left(\xi_{1}+\zeta_{3}\right)\partial^{2}_{\xi_{3}}
+\left(\eta_{2}+\zeta_{3}\right)\partial^{2}_{\eta_{3}} +4
\xi_{2}\left(\partial_{\xi_{1}}+\partial_{\eta_{2}}\right)
 \partial_{\xi_{2}} +4 \xi_{3}\left(\partial_{\xi_{1}}
+\partial_{\zeta_{3}}\right) \partial_{\xi_{3}} \\
&\left.~~~+4\eta_{3}\left(\partial_{\eta_{2}}
+\partial_{\zeta_{3}}\right)
\partial_{\eta_{3}} +2
\eta_{3}\partial_{\xi_{2}}\partial_{\xi_{3}} +2
\xi_{3}\partial_{\xi_{2}}\partial_{\eta_{3}} +2
\xi_{2}\partial_{\xi_{3}}\partial_{\eta_{3}} \right\}
\psi^{\mu \nu \tau}_{q p r}, \end{array} \eqno (27) $$

\noindent
where $\partial_{\xi}$ denotes $\partial/\partial \xi$ and so on.
The second is to apply the Laplace operator to the generalized
spherical harmonic polynomials $Q^{\mu \nu \tau}_{q p r}$.
This part is vanishing because $Q^{\mu \nu \tau}_{q p r}$
satisfies the Laplace equation. The last is the mixed application
$$\begin{array}{l}
2\left\{\left(\partial_{\xi_{1}}\psi^{\mu \nu \tau}_{q p r}
\right)2{\bf x} +\left(\partial_{\xi_{2}}
\psi^{\mu \nu \tau}_{q p r}\right){\bf y}
+\left(\partial_{\xi_{3}}\psi^{\mu \nu \tau}_{q p r}\right)
{\bf z}\right\} \cdot
\bigtriangledown_{\bf x} Q^{\mu \nu \tau}_{q p r}\\
~~~+2\left\{\left(\partial_{\xi_{2}}\psi^{\mu \nu \tau}_{q p r}
\right){\bf x}
+\left(\partial_{\eta_{2}}\psi^{\mu \nu \tau}_{q p r}
\right)2{\bf y} +\left(\partial_{\eta_{3}}
\psi^{\mu \nu \tau}_{q p r}\right){\bf z} \right\} \cdot
\bigtriangledown_{\bf y} Q^{\mu \nu \tau}_{q p r}\\
~~~+2\left\{\left(\partial_{\xi_{3}}\psi^{\mu \nu \tau}_{q p r}
\right){\bf x}
+\left(\partial_{\eta_{3}}\psi^{\mu \nu \tau}_{q p r}\right){\bf y}
+\left(\partial_{\zeta_{3}}\psi^{\mu \nu \tau}_{q p r}\right)
2{\bf z} \right\}\cdot \bigtriangledown_{\bf z} Q^{\mu \nu \tau}_{q p r}.
\end{array} \eqno (28) $$

\noindent
From the definition (14) for $Q^{\mu \nu \tau}_{q p r}$ we have
$$\begin{array}{l}
{\bf x}\cdot \bigtriangledown_{\bf x}Q^{\mu \nu \tau}_{q p r}
=q Q^{\mu \nu \tau}_{q p r},~~~~~~
{\bf y}\cdot \bigtriangledown_{\bf y}Q^{\mu \nu \tau}_{q p r}
=(p+r)Q^{\mu \nu \tau}_{q p r}, \\
{\bf z}\cdot \bigtriangledown_{\bf z}Q^{\mu \nu \tau}_{q p r}
=(\mu+\nu+\tau-q-p-r)Q^{\mu \nu \tau}_{q p r}, \\

{\bf y}\cdot \bigtriangledown_{\bf x}Q^{\mu \nu \tau}_{q p r}
=\left\{ \begin{array}{l}
\displaystyle {(p+1) (q-r) \over q-\nu}Q^{\mu \nu \tau}_{(q-1) (p+1) r}
-\displaystyle {(\mu-q-p+1)(r-\tau+1)\over q-\nu }
Q^{\mu \nu \tau}_{(q-1) p (r+1)},\\
~~~~~~~~~~~~~~~~~~~{\rm when}~~q>\nu, \\
(\nu-q+1)Q^{\mu \nu \tau}_{(q-1) (p+1) r},
~~~~{\rm when}~~q \leq \nu, \end{array} \right.\\

{\bf x}\cdot \bigtriangledown_{\bf y}Q^{\mu \nu \tau}_{q p r}
=\left\{ \begin{array}{l}
(q-\nu+1)Q^{\mu \nu \tau}_{(q+1) (p-1) r},~~~~{\rm when}~~q\geq \nu, \\
\displaystyle {p(q-r+1) \over \nu-q}
Q^{\mu \nu \tau}_{(q+1) (p-1) r}
-\displaystyle {(\mu-q-p+1)(r-\tau+1) \over \nu-q}
Q^{\mu \nu \tau}_{(q+1) (p-2) (r+1)},\\
~~~~~~~~~~~~~~~~~~~ {\rm when}~~q<\nu, \end{array} \right.\\

{\bf z}\cdot \bigtriangledown_{\bf x}Q^{\mu \nu \tau}_{q p r}
=\left\{ \begin{array}{l}
\displaystyle {(\mu-q-p+1) (q-\nu+r-\tau) \over q-\nu}
Q^{\mu \nu \tau}_{(q-1) p r}
-\displaystyle {(p+1)(\nu-r+1)\over q-\nu }\\
~~~~~~~~~~~~~~~~~~~Q^{\mu \nu \tau}_{(q-1) (p+1) (r-1)},
~~~~~~{\rm when}~~q>\nu, \\
-(\nu-q+1)Q^{\mu \nu \tau}_{(q-1) (p+1) (r-1)},~~~~
{\rm when}~~q \leq \nu, \end{array} \right.\\

{\bf x}\cdot \bigtriangledown_{\bf z}Q^{\mu \nu \tau}_{q p r}
=\left\{ \begin{array}{l}
(q-\nu+1)Q^{\mu \nu \tau}_{(q+1) p r},~~~~{\rm when}~~q \geq \nu, \\
\displaystyle {(q+p-\nu+1)(q-r+1) \over \nu-q}
Q^{\mu \nu \tau}_{(q+1) p r}
-\displaystyle {(r-\tau+1)(\mu+\nu-2q-p)\over  \nu-q}\\
~~~~~~~~~~~~~~~\times Q^{\mu \nu \tau}_{(q+1) (p-1) (r+1)},
~~~~{\rm when}~~q < \nu, \end{array} \right.\\

{\bf z}\cdot \bigtriangledown_{\bf y}Q^{\mu \nu \tau}_{q p r}
=\left\{ \begin{array}{l}
(\mu-q-p+1) Q^{\mu \nu \tau}_{q (p-1) r}
+(\nu-r+1) Q^{\mu \nu \tau}_{q p (r-1)}, ~~~~{\rm when}~~q \geq \nu, \\
(\mu-q-p+1)Q^{\mu \nu \tau}_{q (p-1) r}
+(q-r+1) Q^{\mu \nu \tau}_{q p (r-1)},
~~~~{\rm when}~~q < \nu, \end{array} \right.\\

{\bf y}\cdot \bigtriangledown_{\bf z}Q^{\mu \nu \tau}_{q p r}
=\left\{ \begin{array}{l} (p+1) Q^{\mu \nu \tau}_{q (p+1) r}
+(r-\tau+1) Q^{\mu \nu \tau}_{q p (r+1)},~~~~{\rm when}~~q \geq \nu, \\
(q+p-\nu+1) Q^{\mu \nu \tau}_{q (p+1) r}
+(r-\tau+1) Q^{\mu \nu \tau}_{q p (r+1)},~~~~{\rm when}~~q< \nu.
\end{array} \right.
\end{array} \eqno (29) $$

\noindent
Hence, we obtain the generalized radial equation, satisfied
by the functions $\psi^{\mu \nu \tau}_{q p r}(\xi,\eta,\zeta)$:
$$\begin{array}{l}
\bigtriangledown^{2} \psi^{\mu \nu \tau}_{q p r}
+4 q \partial_{\xi_{1}}\psi^{\mu \nu \tau}_{q p r}
+4(p+r) \partial_{\eta_{2}}\psi^{\mu \nu \tau}_{q p r}
+4(\mu+\nu+\tau-p-q-r) \partial_{\zeta_{3}}\psi^{\mu \nu \tau}_{q p r} \\
~~~+\displaystyle {2p(q-r+1) \over q-\nu+1}
\partial_{\xi_{2}}\psi^{\mu \nu \tau}_{(q+1) (p-1) r}
-\displaystyle {2(\mu-q-p)(r-\tau) \over q-\nu+1}
\partial_{\xi_{2}}\psi^{\mu \nu \tau}_{(q+1) p (r-1)} \\
~~~+2(q-\nu) \partial_{\xi_{2}}\psi^{\mu \nu \tau}_{(q-1) (p+1) r}
+\displaystyle {2(\mu-q-p)(q-\nu+r-\tau+1)\over q-\nu+1}
\partial_{\xi_{3}}\psi^{\mu \nu \tau}_{(q+1) p r} \\
~~~-\displaystyle {2 p(\nu-r) \over q-\nu+1}
\partial_{\xi_{3}}\psi^{\mu \nu \tau}_{(q+1) (p-1) (r+1)}
+2(q-\nu) \partial_{\xi_{3}}\psi^{\mu \nu \tau}_{(q-1) p r}
+2(\mu-q-p) \partial_{\eta_{3}}\psi^{\mu \nu \tau}_{q (p+1) r}\\
~~~+2(\nu-r) \partial_{\eta_{3}}\psi^{\mu \nu \tau}_{q p (r+1)}
+2 p \partial_{\eta_{3}}\psi^{\mu \nu \tau}_{q (p-1) r}
+2 (r-\tau) \partial_{\eta_{3}}\psi^{\mu \nu \tau}_{q p (r-1)}\\
=-2\left(E-V\right) \psi^{\mu \nu \tau}_{q p r}, ~~~~~~~{\rm for}~~q> \nu,
\end{array} \eqno (30a) $$
$$\begin{array}{l}
\bigtriangledown^{2} \psi^{\mu \nu \tau}_{q p r}
+4 q \partial_{\xi_{1}}\psi^{\mu \nu \tau}_{q p r}
+4(p+r) \partial_{\eta_{2}}\psi^{\mu \nu \tau}_{q p r}
+4(\mu+\tau-p-r) \partial_{\zeta_{3}}\psi^{\mu \nu \tau}_{q p r} \\
~~~+2p(q-r+1) \partial_{\xi_{2}}\psi^{\mu \nu \tau}_{(q+1) (p-1) r}
-2(\mu-q-p)(r-\tau) \partial_{\xi_{2}}\psi^{\mu \nu \tau}_{(q+1) p (r-1)} \\
~~~+2(p+1)(q-r) \partial_{\xi_{2}}\psi^{\mu \nu \tau}_{(q-1) (p+1) r}
-2(\mu-q-p)(r-\tau)\partial_{\xi_{2}}
\psi^{\mu \nu \tau}_{(q-1) (p+2) (r-1)}\\
~~~+2(\mu-q-p)(r-\tau+1)\partial_{\xi_{3}}\psi^{\mu \nu \tau}_{(q+1) p r}
-2 p(q-r) \partial_{\xi_{3}}\psi^{\mu \nu \tau}_{(q+1) (p-1) (r+1)} \\
~~~+2 p (q-r) \partial_{\xi_{3}}\psi^{\mu \nu \tau}_{(q-1) p r}
-2(r-\tau)(\mu-q-p+1)
\partial_{\xi_{3}}\psi^{\mu \nu \tau}_{(q-1) (p+1) (r-1)}\\
~~~+2(\mu-q-p) \partial_{\eta_{3}}\psi^{\mu \nu \tau}_{q (p+1) r}
+2(q-r) \partial_{\eta_{3}}\psi^{\mu \nu \tau}_{q p (r+1)}
+2 p \partial_{\eta_{3}}\psi^{\mu \nu \tau}_{q (p-1) r}\\
~~~+2 (r-\tau) \partial_{\eta_{3}}\psi^{\mu \nu \tau}_{q p (r-1)}
=-2\left(E-V\right) \psi^{\mu \nu \tau}_{q p r},~~~~~~{\rm for}~~q=\nu,
\end{array} \eqno (30b) $$
$$\begin{array}{l}
\bigtriangledown^{2} \psi^{\mu \nu \tau}_{q p r}
+4 q \partial_{\xi_{1}}\psi^{\mu \nu \tau}_{q p r}
+4(p+r) \partial_{\eta_{2}}\psi^{\mu \nu \tau}_{q p r}
+4(\mu+\nu+\tau-p-q-r) \partial_{\zeta_{3}}\psi^{\mu \nu \tau}_{q p r} \\
~~~+2(\nu-q)\partial_{\xi_{2}}\psi^{\mu \nu \tau}_{(q+1) (p-1) r}
+\displaystyle {2(p+1)(q-r) \over \nu-q+1}
\partial_{\xi_{2}}\psi^{\mu \nu \tau}_{(q-1) (p+1) r} \\
~~~-\displaystyle {2(\mu-q-p)(r-\tau)\over \nu-q+1}
\partial_{\xi_{2}}\psi^{\mu \nu \tau}_{(q-1) (p+2) (r-1)}
-2(\nu-q)\partial_{\xi_{3}}\psi^{\mu \nu \tau}_{(q+1) (p-1) (r+1)} \\
~~~+\displaystyle {2(q+p-\nu)(q-r) \over \nu-q+1}
\partial_{\xi_{3}}\psi^{\mu \nu \tau}_{(q-1) p r}
-\displaystyle {2(r-\tau)(\mu+\nu-2q-p+1) \over \nu-q+1}
\partial_{\xi_{3}}\psi^{\mu \nu \tau}_{(q-1) (p+1) (r-1)}\\
~~~+2(\mu-q-p) \partial_{\eta_{3}}\psi^{\mu \nu \tau}_{q (p+1) r}
+2(q-r) \partial_{\eta_{3}}\psi^{\mu \nu \tau}_{q p (r+1)}
+2 (q+p-\nu) \partial_{\eta_{3}}\psi^{\mu \nu \tau}_{q (p-1) r}\\
~~~+2 (r-\tau) \partial_{\eta_{3}}\psi^{\mu \nu \tau}_{q p (r-1)}
=-2\left(E-V\right) \psi^{\mu \nu \tau}_{q p r}, ~~~~~~{\rm for}~~q<\nu,
\end{array} \eqno (30c) $$

\noindent where $\bigtriangledown^{2} \psi^{\mu \nu \tau}_{q p r}$
is given in Eq. (27). Only six internal variables $\xi_{1}$,
$\xi_{2}$, $\xi_{3}$, $\eta_{2}$, $\eta_{3}$, and $\zeta_{3}$ are
involved both in the equations and in the functions. Eq. (30)
holds either for $D>6$ or for $4 \leq D\leq 6$. For the latter
cases some selfdual representation, antiselfdual representation,
or equivalent representations may occur. Especially, for a
four-body system in $D=4$ dimensions, the representation
$[\mu,1,1]$ is equivalent to the representation $[\mu,0,0]$, but
the generalized radial equations for them are decoupled. They will
be coupled for the $N$-body system with $N>D=4$.

\section{ Quantum $N$-body system in $D$ dimensions }

It is hard to write a unified formulas of the
generalized radial equations for an $N$-body system in
arbitrary $D$-dimensions. However, from the study of the three-body
\cite{gu1,gu2} and four-body system, we are able to summarize
the main features on separating the rotational degrees of
freedom from the internal ones for an $N$-body Schr\"{o}dinger
equation in $D$ dimensions.

First, after removing the motion of the center of mass, there are
$(N-1)$ Jacobi coordinate vectors ${\bf R}_{j}$ for an $N$-body
system. On the other hand, in an $D$-dimensional space it needs
$(D-1)$ vectors to determine the body-fixed frame. When $D\geq N$,
all Jacobi coordinate vectors are used to determine the body-fixed
frame, and all internal variables can be chosen as 
${\bf R}_{j}\cdot {\bf R}_{k}$. The numbers of the rotational variables
and the internal variables are $(N-1)(2D-N)/2$ and $N(N-1)/2$,
respectively. When $D<N$, only $(D-1)$ Jacobi coordinate vectors
are involved to determine the body-fixed frame, and the rest can
be expressed by the first $(D-1)$ Jacobi coordinate vectors
and the internal variables. The set of internal variables
${\bf R}_{j}\cdot {\bf R}_{k}$ is no longer complete because
it could not distinguish two configurations, say with different
${\bf R}_{D}$ reflecting to the superplane spanned by the
first $(D-1)$ Jacobi coordinate vectors. The correct choice
for the internal variables are
$$\begin{array}{l}
\xi_{jk}={\bf R}_{j}\cdot {\bf R}_{k},~~~~~~
\zeta_{\alpha}
=\displaystyle \sum_{a_{1}\ldots a_{D}}~\epsilon_{a_{1}\ldots a_{D}}
R_{1a_{1}}\ldots R_{(D-1)a_{D-1}}R_{\alpha a_{D}}, \\
1\leq j \leq D-1,~~~~~~j\leq k \leq N-1,~~~~~~D\leq \alpha \leq N-1.
\end{array} \eqno (31) $$

\noindent
The numbers of the rotational variables and the internal
variables are $D(D-1)/2$ and $D(2N-D-1)/2$, respectively.

Second, for an $N$-body system in $D$-dimensions ($D\geq N$), 
the angular momentum is described by an irreducible 
representation of SO($D$) denoted by an $(N-1)$-row Young pattern
$[\mu]\equiv [\mu_{1},\mu_{2},\ldots,\mu_{N-1}]$, 
$\mu_{1}\geq \mu_{2} \geq \ldots \geq \mu_{N-1}$. Due to the 
rotational symmetry, one only needs to discuss the eigenfunctions 
of angular momentum with the highest weight. The complete set of
independent base functions with the highest weight consists of 
the eigenfunctions $Q^{[\mu]}_{(q)}({\bf R}_{1},\ldots{\bf R}_{N-1})$
identified by the standard Young tableau $(q)$. Filling the 
digits $1$, $2$, $\ldots$, $N-1$ arbitrarily into a given
Young pattern $[\mu]$ we obtain a young tableau. A Young tableau 
is called standard if the digit in every column of the tableau 
increases downwards and the digit in every row does not decrease 
from left to right. Any standard Young tableau is described by
a set of parameters $(q)$ which contains $(N-1)(N-2)/2$ parameters 
$q_{jk}$, $1\leq k\leq j\leq N-1$, denoting the number of the
digit $j$ in the $k$th row in the standard Young tableau.
The number of independent base functions 
$Q^{[\mu]}_{(q)}({\bf R}_{1},\ldots{\bf R}_{N-1})$ in the 
complete set is equal to the dimension $d_{[\mu]}[SU(N-1)]$
of the irreducible representation $[\mu]$ of the SU($N-1$) 
group. $Q^{[\mu]}_{(q)}({\bf R}_{1},\ldots{\bf R}_{N-1})$ is a homogeneous
polynomial of degree $\sum \mu_{k}$ with respect to the components of
$(N-1)$ Jacobi coordinate vectors ${\bf R}_{j}$, and satisfies the
generalized Laplace equations [see Eq. (20)]. The explicit form
of $Q^{[\mu]}_{(q)}({\bf R}_{1},\ldots{\bf R}_{D-1})$ for the given
standard Young tableau $(q)$ is very easy to write. In the Young
tableau, for each column with the length $t$, filled by digits
$j_{1}<j_{2}<\ldots <j_{t}$,
$Q^{[\mu]}_{(q)}({\bf R}_{1},\ldots{\bf R}_{D-1})$ contains
a determinant as a factor. The $r$th row and $s$th column
in the determinant is $R_{j_{r}(2s-1)}+iR_{j_{r}(2s)}$ if
$D>2(N-1)$. $Q^{[\mu]}_{(q)}({\bf R}_{1},\ldots{\bf R}_{D-1})$
also contains a numerical coefficient for convenience.
When $N\leq D \leq 2(N-1)$, some selfdual representation,
antiselfdual representation and equivalent representations have
to be considered just like the discussion given in the end of Sec. II.
When $D<N$, only the first $(D-1)$ Jacobi coordinate vectors
are involved in the base functions
$Q^{[\mu]}_{(q)}({\bf R}_{1},\ldots{\bf R}_{D-1})$, which are
the same as those for smaller $N=D$.

At last, when $D \geq N$, any wave function $\Psi^{[\mu]}_{\bf M}({\bf
R}_{1},\ldots,{\bf R}_{N-1})$ with the given angular momentum
$[\mu]$ can be expanded with respect to the complete and
independent base functions $Q^{[\mu]}_{(q)}({\bf
R}_{1},\ldots,{\bf R}_{N-1})$
$$\Psi^{[\mu]}_{\bf M}({\bf R}_{1},\ldots,{\bf R}_{N-1})
=\displaystyle \sum_{(q)}~\psi^{[\mu]}_{(q)}(\xi)
Q^{[\mu]}_{(q)}({\bf R}_{1},\ldots,{\bf R}_{N-1}), \eqno (32) $$

\noindent
where the coefficients $\psi^{[\mu]}_{(q)}(\xi)$, called the
generalized radial functions, only depends upon the internal
variables. When $D<N$, $\psi^{[\mu]}_{(q)}(\xi)$ and 
$Q^{[\mu]}_{(q)}({\bf R}_{1},\ldots,{\bf R}_{N-1})$ in Eq. (32)
have to be replaced with $\psi^{[\mu]}_{(q)}(\xi, \zeta)$ and 
$Q^{[\mu]}_{(q)}({\bf R}_{1},\ldots,{\bf R}_{D-1})$, respectively. 
Substituting Eq. (32) into the $N$-body Schr\"{o}dinger
equation in the center-of-mass frame
$$\begin{array}{l}
\displaystyle \sum_{j=1}^{N-1}~\bigtriangledown^{2}_{{\bf R}_{j}}
\Psi^{[\mu]}_{\bf M}({\bf R}_{1},\ldots,{\bf R}_{N-1})
=-2\left\{E-V\left(\xi\right)\right\}
\Psi^{[\mu]}_{\bf M}({\bf R}_{1},\ldots,{\bf R}_{N-1}),
\end{array} \eqno (33) $$

\noindent
one is able to obtain the generalized radial equations. 
The main calculation is to apply the Laplace operator to the
function $\Psi^{[\mu, \nu, \tau]}_{\bf M}({\bf x,y,z})$.
The calculation consists of three parts. The first is to apply
the Laplace operator to the generalized radial functions
$\psi^{[\mu]}_{(q)}(\xi)$, which can be calculated by 
replacement of variables. When $D \geq N$ we have
$$\begin{array}{rl}
\bigtriangledown^{2} \psi^{[\mu]}_{(q)}(\xi)
&=~ \left\{\displaystyle \sum_{j=1}^{N-1}~
\left(4\xi_{jj}\partial^{2}_{\xi_{jj}} 
+2 D\partial_{\xi_{jj}} \right)\right.\\
&~~~+\displaystyle \sum_{j=1}^{N-1}\sum_{k=j+1}^{N-1}~
\left[\left(\xi_{jj}+\xi_{kk}\right)\partial^{2}_{\xi_{jk}}
+4\xi_{jk}\left(\partial_{\xi_{jj}}+\partial_{\xi_{kk}}\right)
\partial_{\xi_{jk}}\right]\\
&\left.~~~+2\displaystyle \sum_{j=1}^{N-1}\sum_{j\neq k=1}^{N-1}
\sum_{j\neq t=k+1}^{N-1}~\xi_{kt}\partial_{\xi_{jk}}\partial_{\xi_{jt}}
\right\}\psi^{[\mu]}_{(q)}(\xi), \end{array} \eqno (34) $$

\noindent
where $\xi_{jk}=\xi_{kj}$ and $\partial_{\xi}$ denotes 
$\partial/\partial \xi$ and so on. The second is to apply 
the Laplace operator to the generalized spherical harmonic 
polynomials. This part is vanishing because the polynomials
satisfy the Laplace equation. The last is the mixed application. 
When $D \geq N$ we have
$$\begin{array}{l}
2\displaystyle \sum_{j=1}^{N-1}~\left\{
\left(\partial_{\xi_{jj}}\psi^{[\mu]}_{(q)}\right)2{\bf R}_{j} 
+\displaystyle \sum_{j\neq k=1}^{N-1}\left(\partial_{\xi_{jk}}
\psi^{[\mu]}_{(q)}\right){\bf R}_{k}\right\} \cdot
\bigtriangledown_{{\bf R}_{j}} Q^{[\mu]}_{(q)}, 
\end{array} \eqno (35) $$

\noindent
where the formulas for 
${\bf R}_{j}\cdot \bigtriangledown_{{\bf R}_{j}}Q^{[\mu]}_{(q)}$
and ${\bf R}_{k}\cdot \bigtriangledown_{{\bf R}_{j}}Q^{[\mu]}_{(q)}$
can be calculated from the property of the polynomial
$Q^{[\mu]}_{(q)}({\bf R}_{1},\ldots{\bf R}_{N-1})$.
When $D<N$, the internal variables have to be chosen as those
given in Eq. (31) so that Eq. (34) becomes more complicated
and Eq. (35) contains more terms of 
$\displaystyle {\partial \zeta_{\alpha}\over \partial {\bf R}_{j}}
\cdot \bigtriangledown_{{\bf R}_{j}}Q^{[\mu]}_{(q)}$ \cite{gu1}.

\section{Conclusions}

In this paper, the problem of separating the rotational
degrees of freedom from the internal ones for the Schr\"{o}dinger
equation of a four-body system in $D$ dimensions is studied
in detail by the method of the generalized spherical harmonic
polynomials. We have found a complete set of independent base
functions with the given angular momentum described by
an irreducible representation $[\mu, \nu, \tau]$ of SO($D$).
This set of base functions have different form for the case
$D>6$ and $3\leq D \leq 6$. We have provided an appropriate
choice of internal variables for this system and derived the
generalized radial equations depending solely on internal
variables. The main features on the problem of separating
the rotational degrees of freedom from the internal ones
for the Schr\"{o}dinger equation of a $N$-body system in
$D$ dimensions is summarized.

\vspace{5mm}
\noindent
{\bf ACKNOWLEDGMENTS} This work is supported by the
National Natural Science Foundation of China.

\end{document}